# Применение скрытой марковской модели для определения PQRST комплексов в электрокардиограммах

Н.С. Шлянкин[1], А.В. Гайдель[1,2]

[1]Самарский национальный исследовательский университет им. академика С.П. Королева, Московское шоссе 34А, Самара, Россия, 443086
[2]Институт систем обработки изображений - филиал ФНИЦ «Кристаллография и фотоника» РАН, Молодогвардейская 151, Самара, Россия, 443001

**Аннотация.** Рассмотрено применение скрытой марковской модели при различных параметрах в задаче сегментирования QRS, ST, T, P, PQ, ISO комплексов электрокардиограмм. Обучение моделей производилось по алгоритму Витерби с помощью базы данных QT Database. Для сравнения был модифицирован алгоритм Пана-Томпкинса поиска продолжительности QRS комплексов.

## 1. Введение

Автоматический анализ электрокардиограмм (ЭКГ) представляет большой интерес на протяжении многих лет в сфере биомедицинской инженерии. Сердечно-сосудистые заболевания по данным Всемирной организации здравоохранения являются одной из самых распространенных причин смерти в мире [1, 2]. ЭКГ является довольно эффективным неинвазивным способом исследования состояния сердечно-сосудистой системы человека, который позволяет получить полезную информацию для определения, диагностики и лечения сердечных заболеваний [3].

Как известно сердце человека имеет высокую электрическую активность, благодаря которой представляется возможным отследить все этапы его сокращения. Электрокардиографы фиксируют биопотенциалы и позволяют проследить за работой сердца. При расшифровке ЭКГ врачи выделяют интервалы и зубцы, каждый из которых имеет свою нормальную продолжительность и высоту [1]. Это позволяет выявить нарушения в работе сердечно-сосудистой системы. При этом наибольший интерес представляют интервалы PQ, ST, RR и зубцы P, QRS, T [1], которые в целом описывают один удар сердца.

Существуют различные методы исследования состояния сердечно-сосудистой системы человека, обзор которых сделан в статье [4]. В настоящее время широкое применение нашли методы анализа ЭКГ на основе различных фильтров и особенностей цифровых сигналов [5, 6, 7]. Однако каждый из этих методов способен сегментировать только один тип зубца или интервала, их точность в некоторых случаях недостаточна.

Задача данной работы – проверить применимость скрытой марковской модели (СММ) для сегментирования ЭКГ сигнала. Работа основана на работах [8, 9]. Рассматривается влияние количества состояний СММ на точность и полноту сегментирования. Преимуществом СММ является способность аннотировать несколько различных комплексов, а не использовать сложную систему из нескольких алгоритмов.

**2. Подготовка датасетов**

Данные для обучения модели были взяты из базы данных QT Database, предоставленной в общий доступ Массачусетским технологическим университетом [10]. База данных содержит 105 пятнадцатиминутных двухканальных отрывков ЭКГ, которые были записаны с частотой 250 Гц. В работе был использован только 1 канал, так как сигналы отличаются своими статистическими параметрами в разных отведениях. Записи были разделены на 95 обучающих записей и 10 записей для тестирования. Для работы с базой данных были использованы инструменты [11].

В базе данных QT Database записи хранятся в виде 2 массивов anntype и annsamp. Anntype содержит название одного из комплексов P, QRS, T, а annsamp время начала и конца этого комплекса. Из этих записей легко выделить промежуточные сегменты. Получается, что между P и QRS лежит интервал PQ, между QRS и T находится ST, а в промежутке между T и P лежит сегмент ISO. Изменив таким образом исходную аннотацию получаем, что вся запись в каждый момент времени содержит один из комплексов ISO, P, PQ, QRS, ST, T, ISO, которые можно увидеть на рисунке 1. Все аннотации были преобразованы к более удобному формату – каждому моменту времени соответствует свой комплекс.

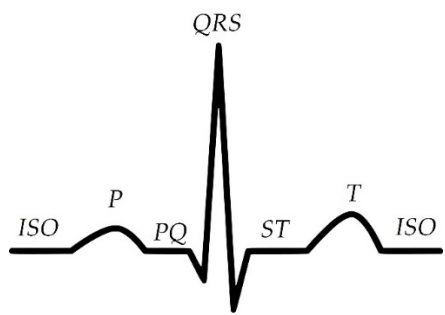

**Рисунок 1.** Удар сердца разбитый на комплексы

**3. Подготовка параметров скрытой марковской модели**

При обработке сигналов ЭКГ часто используют непрерывное вейвлет-преобразование, так как оно характеризует сигнал в различных частотных диапазонах. При этом важно выбрать оптимальный вейвлет, который обеспечивает правильное местоположение комплексов. В случае с СММ вейвлет-преобразование сигнала используется как наблюдаемый параметр модели [8].

Сам сигнал ЭКГ был обработан с помощью вейвлет-преобразования «Мексиканская шляпа», который определяется как

$$Wf(n,j) = \sum_{m=0}^{M-1} f[m] \times \psi_j[m-n],$$

где $f$ – дискретный сигнал длинной $M$, а $\psi_j$ это вейвлет-функция «Мексиканская шляпа» с масштабом $j \in \mathbf{N}$ и $-5 \leq n \leq 5, n \in \mathbf{Z}$:

$$\psi_j = \frac{1}{\sqrt{2^j}} \frac{2}{\sqrt{3}\pi^{1/4}} \left[1 - \left(\frac{n}{2^j}\right)^2\right] \exp\left[\frac{1}{2} \times (n/2^j)^2\right]$$

Из статьи [8] было получено, что основная информация о сигнале содержится при $j = 2, 3, 4$ (так как масштаб $s = 2^j$, то с масштабами 4, 8, 16). Таким образом получилось 3 вейвлет-преобразования сигнала, по которым вычислялись статистические параметры необходимые для обучения модели.

Была построена топология СММ, которую можно посмотреть на рисунке 2. Данная топология учитывает возможное отсутствие комплекса P.

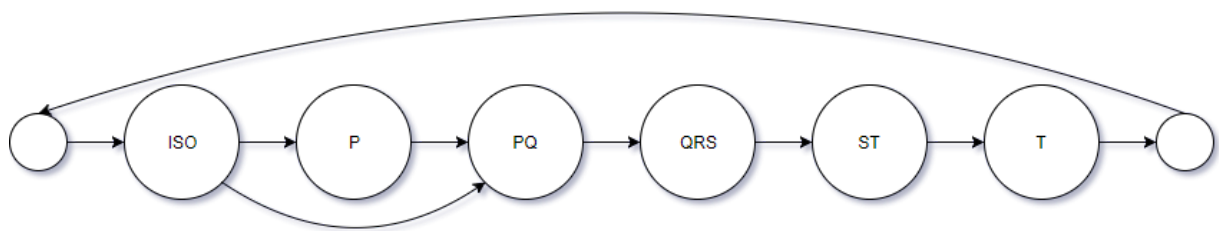

**Рисунок 2.** Топология скрытой марковской модели

В качестве параметра модели задается количество состояний на каждый из комплексов. Для каждого из состояний выделяются фрагменты сигнала и рассчитывается математическое ожидание по каждому из вычисленных ранее вейвлет-преобразований и корреляция между ними.

Формируется матрица переходов состояний, у которой вероятность перехода в следующее состояние равно 1/15, а вероятность остаться в текущем состоянии равна 14/15, при этом учтена вероятность отсутствия P комплекса и сделан переход из состояния соответствующего комплексу ISO в состояние соответствующее комплексу PQ с вероятностью 1/150, а вероятность перехода из ISO в P равна 9/150.

**Таблица 1.** Матрица переходов состояний.

|  | P | PQ | QRS | ST | T | ISO |
|---|---|---|---|---|---|---|
| P | 14/15 | 1/15 | 0 | 0 | 0 | 0 |
| PQ | 0 | 14/15 | 1/15 | 0 | 0 | 0 |
| QRS | 0 | 0 | 14/15 | 1/15 | 0 | 0 |
| ST | 0 | 0 | 0 | 14/15 | 1/15 | 0 |
| T | 0 | 0 | 0 | 0 | 14/15 | 1/15 |
| ISO | 9/150 | 1/150 | 0 | 0 | 0 | 14/15 |

Начальные вероятности равны, так как особой роли при вычислениях они не играют и принято их принимать равными между собой.

## 4. Обучение моделей

Обучение модели проходило по алгоритму Витерби поиска наиболее подходящего списка состояний [12]. Начальными параметрами для обучения модели были подготовленные ранее матрица переходов, начальные вероятности, математическое ожидание, корреляция, количество состояний марковской цепи на каждый из комплексов.

Было обучено 17 моделей с разным количеством состояний на разные комплексы. Каждая из них определяет состояние в каждый момент времени ЭКГ сигнала и выдает последовательность аннотаций. Количество состояний СММ для каждого комплекса можно наблюдать в таблице 2.

**Таблица 2.** Количество состояний на каждый комплекс.

| Комплекс ЭКГ сигнала | Количество состояний моделей | | | | | |
|---|---|---|---|---|---|---|
| QRS | 1 | 3 | 3 | 3 | 3 | 3 |
| P | 1 | 1 | 2 | 3 | 3 | 3 |
| PQ | 1 | 1 | 1 | 2 | 2 | 2 |
| ISO | 1 | 1 | 2 | 3 | 3 | 5 |
| ST | 1 | 1 | 1 | 2 | 2 | 2 |
| T | 1 | 1 | 2 | 3 | 6 | 8 |
| Всего состояний | 6 | 8 | 11 | 16 | 19 | 22 |

Обучающая выборка была разбита на тестовую и обучающую в соотношении 1 к 9 соответственно. Обучение моделей производилось на процессоре Intel Core i5-2500 и видеокарте NVIDIA GeForce GTX 950. Зависимость времени обучения моделей от количества состояний можно наблюдать на рисунке 3.

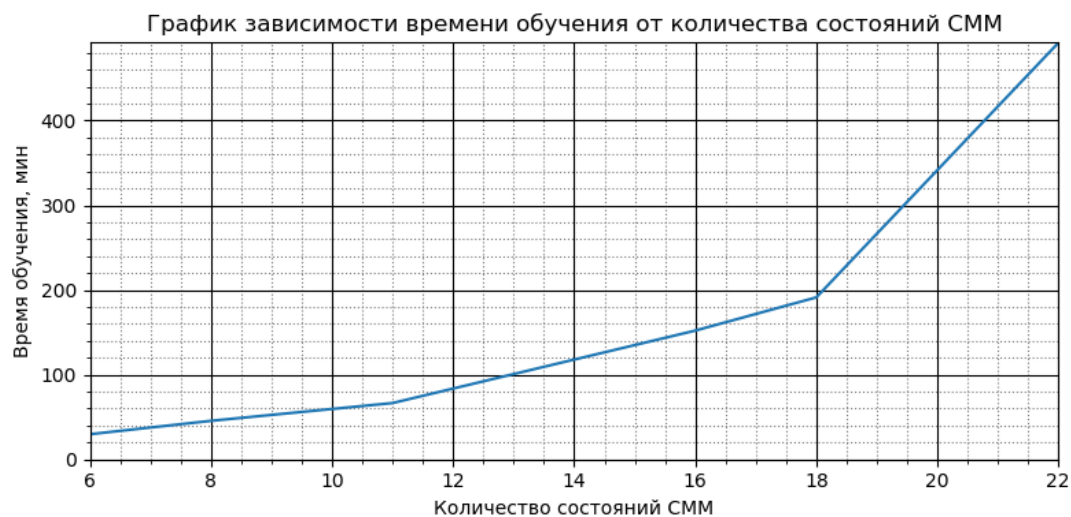

**Рисунок 3.** График зависимости времени обучения от количества состояний скрытой марковской модели

Для каждого из комплексов определялись метрики: $accuracy, precision, recall, fscore$. Метрики моделей можно наблюдать в таблице 3.

**Таблица 3.** Метрики наиболее удачных моделей.

| Комплексы | метрика | Количество состояний моделей | | | | | | Алгоритм Пана-Томпкинса |
|---|---|---|---|---|---|---|---|---|
| | | 6 | 8 | 11 | 16 | 19 | 22 | |
| Для всех предсказываемых комплексов | Accuracy | 0.62 | 0.63 | 0.62 | 0.70 | 0.70 | 0.70 | 0.93 |
| | Precision | 0.54 | 0.56 | 0.59 | 0.62 | 0.64 | 0.63 | - |
| | Recall | 0.61 | 0.67 | 0.61 | 0.68 | 0.67 | 0.66 | |
| | Fscore | 0.55 | 0.58 | 0.55 | 0.64 | 0.62 | 0.62 | |
| QRS | Precision | 0.63 | 0.58 | 0.70 | 0.74 | 0.81 | 0.82 | 0.87 |
| | Recall | 0.91 | 0.95 | 0.87 | 0.84 | 0.84 | 0.83 | 0.55 |
| | Fscore | 0.75 | 0.72 | 0.77 | 0.79 | 0.82 | 0.83 | 0.68 |
| PQ | Precision | 0.31 | 0.38 | 0.41 | 0.35 | 0.35 | 0.31 | |
| | Recall | 0.49 | 0.68 | 0.44 | 0.47 | 0.48 | 0.51 | |
| | Fscore | 0.38 | 0.49 | 0.42 | 0.40 | 0.41 | 0.39 | |
| P | Precision | 0.30 | 0.36 | 0.26 | 0.40 | 0.36 | 0.36 | |
| | Recall | 0.66 | 0.78 | 0.78 | 0.76 | 0.78 | 0.67 | |
| | Fscore | 0.41 | 0.49 | 0.39 | 0.52 | 0.50 | 0.47 | – |
| ISO | Precision | 0.83 | 0.84 | 0.76 | 0.81 | 0.88 | 0.87 | |
| | Recall | 0.63 | 0.57 | 0.57 | 0.71 | 0.65 | 0.69 | |
| | Fscore | 0.71 | 0.68 | 0.65 | 0.76 | 0.75 | 0.77 | |
| ST | Precision | 0.44 | 0.49 | 0.74 | 0.69 | 0.76 | 0.77 | |
| | Recall | 0.44 | 0.41 | 0.25 | 0.59 | 0.39 | 0.38 | |
| | Fscore | 0.44 | 0.44 | 0.37 | 0.64 | 0.51 | 0.51 | |
| T | Precision | 0.73 | 0.74 | 0.66 | 0.74 | 0.67 | 0.68 | |
| | Recall | 0.57 | 0.63 | 0.73 | 0.70 | 0.88 | 0.85 | |
| | Fscore | 0.64 | 0.68 | 0.69 | 0.72 | 0.76 | 0.76 | |

Из таблицы видно, что в зависимости от количества состояний на комплекс метрики моделей меняются незначительно. Наиболее интересной метрикой является $fscore$, которая является среднегармоническим метрик $precision$ и $recall$, так как с помощью нее легко оценить какая из моделей показывает лучшие результаты. Метрика $accuracy$ оценивает процент правильно предсказанных моментов времени. Модель с 16 состояниями показывает результат немного лучше моделей с меньшим и большим количеством состояния.

Сравним работу СММ с алгоритмом Пана-Томпкинса поиска QRS комплексов из статьи [13]. В алгоритм был добавлен дополнительный этап, который определит продолжительность QRS комплексов, так как стандартный алгоритм предназначен только для обнаружения местоположения, а не продолжительности комплекса. Берутся три последовательно идущих пика $R1, R2, R3$. На отрезке $\left[\frac{R1+R2}{2}; \frac{R2+R3}{2}\right]$ лежит QRS комплекс, которому соответствует зубец $R2$. На этом отрезке определяем порог:

$$threshold = average\left(x_p\left[\frac{R1+R2}{2}:\frac{R2+R3}{2}\right]\right) + gamma \cdot \max\left(x_p\left[\frac{R1+R2}{2}:\frac{R2+R3}{2}\right]\right),$$

где $average\left(x_p\left[\frac{R1+R2}{2}:\frac{R2+R3}{2}\right]\right)$ – среднее значение сигнала на интервале, $gamma$ – коэффициент меньше 1, $\max\left(x_p\left[\frac{R1+R2}{2}:\frac{R2+R3}{2}\right]\right)$ – максимум на интервале, $gamma$ – множитель позволяющий изменять величину порога в зависимости от пологости QRS сегментов обработанного сигнала, $x_p$ – обработанный по алгоритму Пана-Томпкинса сигнал. Находим начало и конец сегмента, значение сигнала на котором в пределах порога и определяем его как QRS комплекс.

Модифицированный алгоритм неплохо определяет сегменты QRS комплекса, однако СММ показывает результаты лучше, если оценивать по метрике $fscore$. Конечно, можно создать более совершенную модификацию алгоритма Пана-Томпкинса, которая будет обходить СММ в предсказании QRS комплекса, при этом данный метод не требует предварительного обучения, однако алгоритм уже достаточно сложен в реализации, производит большое количество операций, при этом определяет лишь один сегмент, а СММ способна сегментировать сразу несколько комплексов.

**5. Заключение**

Было обучено несколько скрытых марковских моделей с различным числом состояний. Практика показала, что изменение количества вплоть до 16 состояний дает увеличение точности сегментирования до 70% и полноты до 68%, однако увеличивается и время обучения модели. Если сравнивать с классическим алгоритмом Пана-Томпкинса, то можно сказать о том, что СММ лучше применять если необходимо определить продолжительность нескольких комплексов.

СММ дает относительно неплохие результаты при сегментировании сигналов, однако для большего увеличения точности предсказания необходим другой подход. Дальнейшие исследования могут быть направлены на обучение СММ на ЭКГ сигналах одного человека, что позволит более точно определить статистические параметры, так как они будут более стабильны.

**6. Литература**

# Application of the Hidden Markov Model for determining PQRST complexes in electrocardiograms

**N.S. Shlyankin[1], A.V. Gaidel[1,2]**

[1] Samara University, Moskovskoe Shosse 34A, Samara, Russia, 443086
[2] IPSI RAS - branch of the FSRC «Crystallography and Photonics» RAS, Molodogvardejskaya street 151, Samara, Russia, 443001

**Abstract.** The application of the hidden Markov model with various parameters in the segmentation task of QRS, ST, T, P, PQ, ISO complexes of electrocardiograms is considered. Models were trained using the Viterbi algorithm using the QT Database. For comparison, the Pan-Tompkins algorithm for searching for the duration of QRS complexes was modified.